\documentclass[10pt,twocolumn]{IEEEtran}
\usepackage{}
\usepackage[latin1]{inputenc}
\usepackage{amssymb}
\usepackage{latexsym}
\usepackage{mathrsfs}
\usepackage{amsfonts}
\usepackage{amsbsy}
\usepackage{amsmath}
\usepackage{times}
\usepackage{epsfig,epsf,times,amssymb,amsmath}
\usepackage{color}
\usepackage{setspace}

\newcommand{\argmin}{\operatornamewithlimits{argmin}}
\newcommand{\argmax}{\operatornamewithlimits{argmax}}

\IEEEoverridecommandlockouts
\begin{document}
\title{Study of Buffer-Aided Distributed Space-Time Coding for Cooperative Wireless Networks}

\author{\IEEEauthorblockN{Tong Peng and Rodrigo C. de Lamare} \vspace{-2em}
\thanks{CETUC/PUC-RIO, BRAZIL, Communications Research Group. Email: tong.peng@cetuc.puc-rio.br; delamare@cetuc.puc-rio.br}
\thanks{This research is supported by the National Council for Scientific and Technological Development (CNPq) in Brazil.}}

\maketitle

\IEEEpeerreviewmaketitle

\begin{abstract}
 {This work proposes adaptive buffer-aided distributed space-time
coding schemes and algorithms with feedback for wireless networks equipped with
buffer-aided relays. The proposed schemes employ a maximum likelihood receiver
at the destination and adjustable codes subject to a power constraint with an
amplify-and-forward cooperative strategy at the relays. Each relay is equipped
with a buffer and is capable of storing blocks of received symbols and
forwarding the data to the destination if selected. Different antenna
configurations and wireless channels, such as static block fading channels, are
considered. The effects of using buffer-aided relays to improve the bit error
rate (BER) performance are also studied. Adjustable relay selection and
optimization algorithms that exploit the extra degrees of freedom of relays
equipped with buffers are developed to improve the BER performance. We also
analyze the pairwise error probability and diversity of the system when using
the proposed schemes and algorithms in a cooperative network. Simulation
results show that the proposed schemes and algorithms obtain performance gains
over previously reported techniques.}
\end{abstract}

\section{Introduction}

Cooperative relaying systems, which employ relay nodes with an arbitrary number
of antennas between the source node and the destination node as a distributed
antenna array, can obtain diversity gains by employing space-time coding (STC)
schemes to improve the reliability of wireless links
\cite{J.N.Laneman2004,Clarke1}. In existing cooperative relaying systems,
amplify-and-forward (AF), decode-and-forward (DF) or compress-and-forward (CF)
\cite{J.N.Laneman2004} cooperation strategies are often employed with the help
of multiple relay nodes.

The adoption of distributed space-time coding (DSTC) schemes at relay nodes in
a cooperative network, providing more copies of the desired symbols at the
destination node, can offer the system diversity and coding gains which enable
more effective interference mitigation and enhanced performance. A recent focus
of DSTC techniques lies in the design of full-diversity schemes with minimum
outage probability \cite{Maham}-\cite{Maham Birsen}. In \cite{Maham}, the GABBA
STC scheme has been extended to a distributed multiple-input and
multiple-output (MIMO) network with full-diversity and full-rate, while an
optimal algorithm for the design of DSTC schemes that achieve the optimal
diversity and multiplexing tradeoff has been derived in \cite{Sheng}. A
quasi-orthogonal distributed space-time block coding (DSTBC) scheme for
cooperative MIMO networks is presented and shown to achieve full rate and full
diversity with any number of antennas in \cite{Maham Birsen}. In
\cite{B.Sirkeci-Mergen}, an STC scheme that multiplies a randomized matrix by
the STC code matrix at the relay node before the transmission is derived and
analyzed. The randomized space-time coding (RSTC) schemes can achieve the
performance of a centralized STC scheme in terms of coding gain and diversity
order.

Relay selection algorithms such as those designed in
\cite{Clarke1,Clarke2} provide an efficient way to assist the
communication between the source node and the destination node.
Although the best relay node can be selected according to different
optimization criteria, conventional relay selection algorithms often
focus on the best relay selection (BRS) scheme \cite{A.Bletsas},
which selects the links with maximum instantaneous signal-to-noise
ratio ($SNR$). The best relay forwards the information to the
destination which results in an improved BER performance. Recently,
cooperative schemes with more general configurations involving a
source node, a destination node and multiple relays equipped with
buffers has been introduced and analyzed in
\cite{N.Zlatanov}-\cite{abaro}. The main idea is to select the best
link during each time slot according to different criteria, such as
maximum instantaneous $SNR$ and maximum throughput. In
\cite{N.Zlatanov}, an introduction to buffer-aided relaying networks
is given, and further analysis of the throughput and diversity gain
is provided in \cite{N.Zlatanov2}. In \cite{N.Zlatanov3} and
\cite{N.Zlatanov4}, an adaptive link selection protocol with
buffer-aided relays is proposed and an analysis of the network
throughput and the outage probability is developed. A max-link relay
selection scheme focusing on achieving full diversity gain, which
selects the strongest link in each time slot is proposed in
\cite{I.Krikidis}. A max-max relay selection algorithm is proposed
in \cite{A.Ikhlef2} and has been extended to mimic a full-duplex
relaying scheme in \cite{A.Ikhlef} with the help of buffer-aided
relays.

Despite the early work with buffer-aided relays and its performance
advantages, schemes that employ STC techniques have not been
considered so far. In particular, STC and DSTC schemes encoded at
the relays can provide higher diversity order and higher reliability
for wireless systems.  {In this work, we propose adjustable
buffer-aided distributed and non-distributed STC schemes, relay
selection and adaptive buffer-aided relaying optimization (ABARO)
algorithms for cooperative relaying systems with feedback.} We
examine two basic configurations of relays with STC and DSTC
schemes: one in which the coding is performed independently at the
relays \cite{B.Sirkeci-Mergen}, denoted multiple-antenna system
(MAS) configuration, and another in which coding is performed across
the relays \cite{Maham Birsen}, called single-antenna system (SAS)
configuration.  {According to the literature, STC schemes can be
implemented at a single relay node with multiple antennas and DSTC
schemes can be used at multiple relay nodes with a single antenna.
Moreover, an adjustable STC scheme is developed in \cite{TARMO}
which indicates that by using an adjustable coding vector at
single-antenna relay nodes, a complete STC scheme can be
implemented. In this work, we consider a STC scheme implemented at a
multiple-antenna relay node and a DSTC scheme applied at a group of
single-antenna relay nodes along with adjustable STC and DSTC
schemes at both types of relays.}  {Compared to relays without
buffers, buffer-aided relays help mitigate deep fading periods
during communication between devices as the received symbols can be
stored at the relays, which contributes to a significant BER
performance improvement. Although the delay is a key issue for
buffer-aided relays, their key advantage is to improve the error
tolerance and transmission accuracy of the links in the network.
Buffer-aided relay schemes can be used in networks in which the
delay is not an issue and with delay tolerance.}

The proposed schemes, relay selection and ABARO optimization
algorithms can be structured into two parts, the first one is the
relay selection part which chooses the best link with the maximum
instantaneous $SNR$ or signal to interference and noise ratio
($SINR$) and checks if the state of the best relay node is available
to transmit or receive, and the second part refers to the
optimization of the adjustable STC schemes employed at the relay
nodes. ABARO is based on the maximum-likelihood (ML) criterion
subject to constraints on the transmitted power at the relays for
different cooperative systems. STC schemes are employed at each
relay node and an ML detector is employed at the destination node in
order to ensure full receive diversity. Suboptimal receive and
beamforming approaches
\cite{mmimo,wence,Costa,delamare_ieeproc,TDS_clarke,TDS_2,armo,buffer,switch_int,switch_mc,smce,TongW,jpais_iet,TARMO,keke1,kekecl,keke2,Tomlinson,dopeg_cl,peg_bf_iswcs,gqcpeg,peg_bf_cl,Harashima,mbthpc,zuthp,rmbthp,rprec,Hochwald,BDVP},
\cite{delamare_mber,rontogiannis,delamare_itic,stspadf,choi,stbcccm,FL11,jio_mimo,peng_twc,spa,spa2,memd,jio_mimo,P.Li,jingjing,did,bfidd,mbdf,mbermmimo}.
 and advanced signal processing techniques
\cite{scharf,bar-ness,pados99,reed98,hua,goldstein,santos,qian,delamarespl07,xutsa,delamaretsp,kwak,xu&liu,delamareccm,wcccm,delamareelb,jidf,delamarecl,delamaresp,delamaretvt,jioel,delamarespl07,delamare_ccmmswf,jidf_echo,delamaretvt10,delamaretvt2011ST,delamare10,fa10,lei09,ccmavf,lei10,jio_ccm,ccmavf,stap_jio,zhaocheng,zhaocheng2,arh_eusipco,arh_taes,dfjio,rdrab,dcg_conf,dcg,dce,drr_conf,dta_conf1,dta_conf2,dta_ls,song,wljio,barc,jiomber,saalt,alrdoa}.
can be also used at the destination node to reduce the detection
complexity. Moreover, stochastic gradient (SG) adaptive algorithms
\cite{S.Haykin} are developed in order to compute the required
parameters at a reduced computational complexity. We study how the
adjustable codes can be employed at buffer-aided relays combined
with relay selection and how to optimize the adjustable codes by
employing an ML criterion. {A feedback channel is required in the
proposed scheme and algorithms. All the computations are done at the
destination node so that the useful information, such as relay
selection information and optimized coding matrices are assumed
known. We have studied the impact of feedback errors in
\cite{TARMO}, however, in this work we focus on the effects of using
the proposed buffer-aided relay schemes, relay selection and
optimization algorithms. The feedback is assumed to be error-free
and the devices are assumed to have perfect channel state
information (CSI).} The proposed relay selection and optimization
algorithms can be implemented with different types of STC and DSTC
schemes in cooperative relaying systems with DF or AF protocols. We
first study the design of adjustable STC schemes and relay selection
algorithms for single-antenna systems and then extend it to
multiple-antenna systems, which enable further diversity gains or
multiplexing gains. The proposed algorithms and schemes are also
considered with DSTC schemes. In single-antenna networks, DSTC
schemes are used with an arbitrary number of relays and a group of
relays is selected to implement the DSTC scheme. In multiple-antenna
networks, a complete DSTC scheme can be obtained at each relay node
and a superposition of multiple DSTC transmissions is received at
the destination.

This paper is organized as follows. Section II introduces a cooperative two-hop
relaying systems with multiple buffer-aided relays applying the AF strategy in
SAS and MAS configurations, respectively. In Section III the detailed
adjustable STC scheme is introduced. The proposed relay selection and code
optimization algorithms are derived in Section IV and the DSTC schemes are
considered in Section V. The analysis of the proposed algorithms is shown in
Section VI, whereas in Section VII we provide the simulation results. Section
VIII gives the conclusions of the work.

Notation: the italic, the bold lower-case and the bold upper-case letters
denote scalars, vectors and matrices, respectively. The operator
$\parallel{\boldsymbol X}\parallel_F=\sqrt{{\rm Tr}({\boldsymbol
X}^H\cdot{\boldsymbol X})}=\sqrt{{\rm Tr}({\boldsymbol X}\cdot{\boldsymbol
X}^\emph{H})}$ is the Frobenius norm. ${\rm Tr}(\cdot)$ stands for the trace of
a matrix, and the $N \times N$ identity matrix is written as ${\boldsymbol
I}_N$.

\section{Cooperative System Models}

In this section, we introduce the cooperative system models adopted
to evaluate the proposed schemes and algorithms. We consider two
relay configurations: SAS in which each node contains only a single
antenna and MAS in which each node contains multiple antennas.
{This work focuses on the relay selection and adjustable code
matrices optimization algorithms so that we assume that perfect CSI
is available at the relays and destination nodes. However, we remark
that that CSI can be obtained in practice by using pilot sequences
and cooperative channel estimation algorithms \cite{smce,segce}. }

\subsection{Cooperative System Models for SAS}

\begin{figure}
\begin{center}
\def\epsfsize#1#2{0.825\columnwidth}
\epsfbox{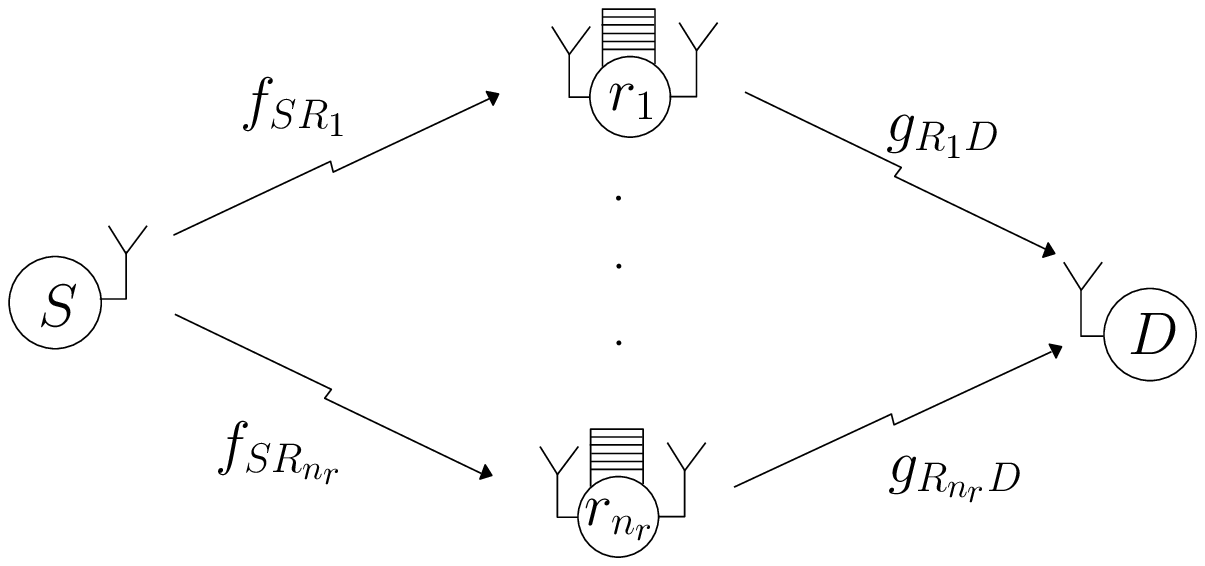}
\caption{Cooperative System Model with $n_r$ Relay Nodes}\label{f1}
\vspace{-1em}
\end{center}
\end{figure}

In this section, we consider a two-hop system that is shown in Fig.1 and which
consists of a source node, a destination node and $n_r$ relays. Each node
contains a single antenna. Let ${\boldsymbol s}[j]$ denote a block of modulated
data symbols with length of $M$ and covariance matrix $E\big[{\boldsymbol
s}[j]{\boldsymbol s}^\emph{H}[j]\big] = \sigma_{s}^{2}{\boldsymbol I}_M$, where
$\sigma_s^2$ denotes the signal power and $j$ is the index of the blocks. We
assume that the channels are static over the transmission period of
${\boldsymbol s}[j]$. The minimum buffer size is equal to the size of one block
of symbols, $M$, and the maximum buffer size is equal to $MJ$, where $J$ is the
maximum number of symbol blocks. In the first hop, the source node sends the
modulated symbol vector $\boldsymbol{s}[j]$ to the relay nodes and the received
data are given by
\begin{equation}\label{2.1.1}
\begin{aligned}
    \boldsymbol{r}_{SR_k}[j] &= \sqrt{P_S}f_{SR_k}[j]\boldsymbol{s}[j] + \boldsymbol{n}_{SR_k}[j], ~k=1,2,...,n_r, ~j = 1,2,...J,
\end{aligned}
\end{equation}
where $f_{SR_k}[j]$ denotes the channel state information (CSI) between the
source node and the $k$th relay, and $\boldsymbol{n}_{SR_k}[j]$ stands for the
$M \times 1$ additive white Gaussian noise (AWGN) vector generated at the $k$th
relay with variance $\sigma^2_r$. The transmission power assigned at the source
node is denoted as $P_S$. At the relay nodes, in order to implement an STC
scheme the received symbols are divided into $i=M/N$ groups, where $N$ denotes
the number of symbols required to encode an STC scheme and whose value is
different according to different STC schemes, e.g. $N=2$ for the $2 \times 2$
Alamouti STBC scheme and $N=4$ for the linear dispersion code (LDC) scheme in
\cite{Hassibi_LDC}. The transmission in the second hop is expressed as follows:
\begin{equation}\label{2.1.2}
\begin{aligned}
    \boldsymbol{r}_{R_kD}[i] &= \sqrt{P_R}g_{R_kD}[i]{\boldsymbol c}_{rand}[i] + \boldsymbol{n}_{R_kD}[i], ~k=1,2,...,n_r, ~i = 1,2,...,M/N,
\end{aligned}
\end{equation}
where $\boldsymbol{r}_{R_kD}[i]$ denotes the $i$th $T \times 1$ received symbol
vector. The $T \times 1$ adjustable STC scheme is denoted by ${\boldsymbol
c}_{rand}[i]$, and $g_{R_kD}[i]$ denotes the CSI factor between the $k$th relay
and the destination node. The transmission power assigned at the relay node is
denoted as $P_R$. The vector $\boldsymbol{n}_{R_kD}[i]$ stands for the AWGN
vector generated at the destination node with variance $\sigma^2_d$. It is
worth mentioning that during the transmission period of each group the channel
is static. The details of adjustable STC encoding and decoding procedures are
given in the next section.

\subsection{Cooperative System Models for MAS}

\begin{figure}
\begin{center}
\def\epsfsize#1#2{0.825\columnwidth}
\epsfbox{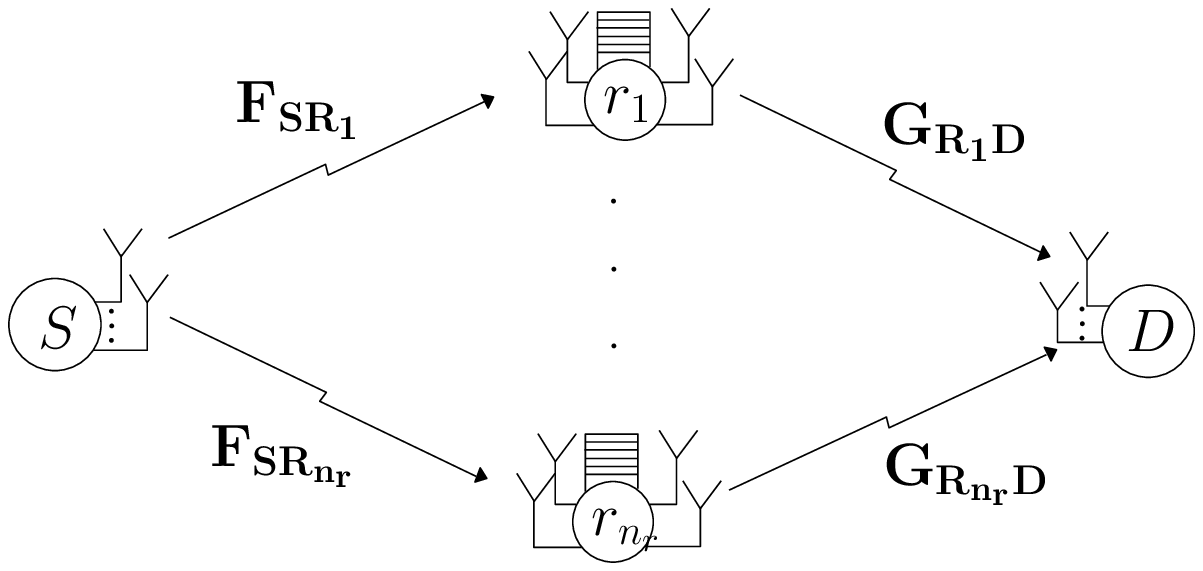}
\caption{Cooperative System Model with $n_r$ Relay Nodes}\label{f1}
\vspace{-1em}
\end{center}
\end{figure}

In this section, we extend the single-antenna system model to a two-hop
multiple-antenna system that is shown in Fig.2. Each node contains $N \geq 2$
antennas. Let ${\boldsymbol s}[j]$ denote a modulated data symbol vector with
length $M$, which is a block of symbols in a packet. The data symbol vector
${\boldsymbol s}[j]$ can be sent from the source to the relays within one time
slot since multiple antennas are employed. We assume that the channels are
static over the transmission period of ${\boldsymbol s}[j]$ and, for
simplicity, we assume that $N = M$ and the minimum buffer size is equal to $M$.
In the first hop, the source node sends ${\boldsymbol s}[j]$ to the relay nodes
and the received data are described by
\begin{equation}\label{3.1.1}
\begin{aligned}
    \boldsymbol{r}_{SR_k}[j] &= \sqrt{\frac{P_S}{N}} \boldsymbol{F}_{SR_k}\boldsymbol{s}[j]
    + \boldsymbol{n}_{SR_k}[j], ~k=1,2,...,n_r, ~j = 1,2,...J,
\end{aligned}
\end{equation}
where $\boldsymbol{F}_{SR_k}[j]$ denotes the $N \times N$ CSI matrix between
the source node and the $k$th relay, and $\boldsymbol{n}_{SR_k}[j]$ stands for
the $N \times 1$ AWGN vector generated at the $k$th relay with variance
$\sigma^2_r$. At each relay node, an adjustable code vector is randomly
generated before the forwarding procedure and the received data are expressed
as
\begin{equation}\label{3.1.3}
\begin{aligned}
  & {\boldsymbol R}_{R_kD}[j] =
  \sqrt{\frac{P_R}{N}}\boldsymbol{G}_{R_kD}[j]{\boldsymbol V}[j]{\boldsymbol C}[j]
  + {\boldsymbol N}_{R_kD}[j]=\sqrt{\frac{P_R}{N}}\boldsymbol{G}_{R_kD}[j]{\boldsymbol C}_{rand}[j]
  + {\boldsymbol N}_{R_kD}[j],\\ &~~~~~~~~~~~~~~~~~~~~~~~~~~~k=1,2,...,n_r, ~j = 1,2,...J,
\end{aligned}
\end{equation}
where ${\boldsymbol C}[j]$ denotes the $N \times T$ standard STC scheme and
${\boldsymbol V}[j]={\rm diag} \{{\boldsymbol v}[j]\}$ stands for the $N \times
N$ diagonal adjustable code matrix whose elements are from the adjustable
vector ${\boldsymbol v}=[v_1, v_2,...,v_N]$. The $N \times T$ adjustable code
matrix is denoted by ${\boldsymbol C}_{rand}[j]$. An equivalent representation
of the received data is given by the received vector ${\boldsymbol
r}_{R_kD}[j]$, which replaces the received symbol matrix ${\boldsymbol
R}_{R_kD}[j]$ in ({\ref{3.1.3}}) and is written as
\begin{equation}\label{3.1.4}
\begin{aligned}
    {\boldsymbol r}_{R_kD}[j] = & \sqrt{\frac{P_R P_S}{N}}{\boldsymbol V}_{eq}[j]
    {\boldsymbol H}[j]{\boldsymbol s}[j] + \sqrt{\frac{P_R}{N}}{\boldsymbol V}_{eq}[j]
    {\boldsymbol G}_{R_kD}[j]{\boldsymbol n}_{sr_k}[j] + {\boldsymbol n}_{R_kD}[j]\\
    =& \sqrt{\frac{P_R P_S}{N}}{\boldsymbol V}_{eq}[j]{\boldsymbol H}[j]{\boldsymbol s}[j] + {\boldsymbol n}[j],
\end{aligned}
\end{equation}
where ${\boldsymbol V}_{eq}[j]={\boldsymbol I}_{T \times T} \otimes
{\boldsymbol V}[j]$ denotes the $TN \times TN$ block diagonal equivalent
adjustable code matrix and $\otimes$ is the Kronecker product, and
${\boldsymbol H}[j]$ stands for the equivalent channel matrix which is the
combination of ${\boldsymbol F}_{SR_k}[j]$ and ${\boldsymbol G}_{R_kD}[j]$. The
$TN \times 1$ vector ${\boldsymbol n}[i]$ contains the equivalent received
noise vector at the destination node, which can be modeled as an AWGN with zero
mean and covariance matrix $(\sigma^{2}_d+\|{\boldsymbol V}_{eq}[j]{\boldsymbol
G}_{R_kD}[j]\|^2_F\sigma^{2}_r){\boldsymbol I}_{NT}$.

\section{Adjustable Space-Time Coding Scheme}

In this section, we detail the adjustable STC schemes in the SAS and MAS
configurations. The encoding procedure of the adjustable coding schemes as
compared to standard STC and DSTC schemes is different in the SAS and the MAS
configuration, and we describe them in the following.

\subsection{Adjustable Space-Time Coding Scheme for SAS}

Here, we develop the procedure of adjustable STC for the SAS configuration. In
\cite{B.Sirkeci-Mergen} and \cite{TARMO}, adjustable codes are employed to
allow relays with a single antenna to transmit STC schemes. In the second hop,
the whole packet will be forwarded to the destination node. Due to the
consideration of the performance of an $N \times T$ STC scheme, the received
packet is divided into $i=M/N$ groups and each group contains $N$ symbols.
These $N$ symbols will be encoded by an STC generation matrix and then
forwarded to the destination. For example, suppose that a packet contains
$M=100$ symbols and the $2 \times 2$ Alamouti space-time block coding (STBC)
scheme is used at the relay nodes. We first split ${\boldsymbol r}_{SR_k}$ into
$50$ groups, encode the symbols in the first group by the Alamouti STBC scheme
and then multiply a $1 \times 2$ randomized vector ${\boldsymbol v}$. The
original $2 \times 2$ orthogonal Alamouti STBC scheme ${\boldsymbol C}$ results
in the following code:
\begin{equation}\label{2.1.5}
\begin{aligned}
    {\boldsymbol c}_{rand} &= {\boldsymbol v}{\boldsymbol C} = \left[v_1~ v_2\right]\left[\begin{array}{cc} r_{SR_k}1 & -r^*_{SR_k}2 \\ r_{SR_k}2 & ~r^*_{SR_k}1 \end{array} \right]\\ & = \left[v_1r_{SR_k}1+v_2r_{SR_k}2 ~~ v_2r^*_{SR_k}1-v_1r^*_{SR_k}2\right],
\end{aligned}
\end{equation}
where $r_{SR_k}1$ and $r_{SR_k}2$ are symbols in the first group, and the $1
\times 2$ vector ${\boldsymbol v}$ denotes the randomized vector whose elements
are generated randomly according to different criteria described in
\cite{B.Sirkeci-Mergen}. As shown in (\ref{2.1.5}), the $2 \times 2$ STBC
matrix changes to a $1 \times 2$ STBC vector which can be transmitted by a
relay node with a single antenna in $2$ time slots. Different STC schemes such
as the LDC scheme in \cite{Hassibi_LDC} can be easily adapted to the randomized
vector encoding in (\ref{2.1.5}). Therefore, the transmission of the randomized
STC schemes can be described as
\begin{equation}\label{2.1.6}
    {\boldsymbol r} = \sqrt{P_T}h{\boldsymbol c}_{rand} + {\boldsymbol n} = \sqrt{P_T}h{\boldsymbol v}{\boldsymbol C} + {\boldsymbol n},
\end{equation}
where $h$ denotes the channel coefficient which is assumed to be constant
within the transmission time slots, and ${\boldsymbol n}$ stands for the noise
vector. The decoding methods of the randomized STC schemes are the same as that
of the original STC schemes. At the destination, instead of the estimation of
the channel coefficient $h$, the resulting composite parameter vector
${\boldsymbol v}h$ is estimated. As a result, the transmission of a randomized
STC vector is similar to the transmission of a deterministic STC scheme over an
effective channel. Taking the randomized Alamouti scheme as an example, the
linear ML decoding for the information symbols $s_1$ and $s_2$ is given by
\begin{equation}\label{2.1.6.1}
    \tilde{s_1} = h^*_{rand1}r_1 + h_{rand2}r^*_2, ~\tilde{s_2} = h^*_{rand2}r^*_1 + h_{rand1}r^*_2,
\end{equation}
where $h_{rand1}$ and $h_{rand2}$ are the randomized channel coefficients in
${\boldsymbol v}h$. Different decoding methods can be employed in this context.
In \cite{TARMO}, optimization algorithms to compute the randomized code vector
${\boldsymbol v}$ are proposed in order to obtain a performance improvement.

Since the adjustable STC scheme is employed at the relay node, the received
vector ${\boldsymbol r}_{R_kD}[i]$ in ({\ref{2.1.2}}) can be rewritten as
\begin{equation}\label{2.1.4}
\begin{aligned}
    {\boldsymbol r}_{R_kD}[i] =& \sqrt{P_R P_S}{\boldsymbol V}_{eq}[i]h[i]{\boldsymbol s}[i] + \sqrt{P_R}{\boldsymbol V}_{eq}[i]g_{R_kD}[i]{\boldsymbol n}_{sr_k}[i] + {\boldsymbol n}_{R_kD}[i]\\ =& \sqrt{P_R P_S}{\boldsymbol V}_{eq}[i]h[i]{\boldsymbol s}[i] + {\boldsymbol n}[i],
\end{aligned}
\end{equation}
where ${\boldsymbol V}_{eq}[i]$ denotes the $T \times N$ block diagonal
equivalent adjustable code matrix, and $h[i]=f_{SR_k}[i]g_{R_kD}[i]$ stands for
the equivalent channel. The vector ${\boldsymbol n}[i]$ contains the equivalent
received noise vector at the destination node, which can be modeled as an AWGN
with zero mean and covariance matrix $(\sigma^{2}_d+\|{\boldsymbol
V}_{eq}[i]g_{R_kD}[i]\|^2_F\sigma^{2}_r){\boldsymbol I}_{NT}$.

\subsection{Adjustable Space-Time Coding Scheme for MAS}

In this section, the details of the adjustable STC encoding procedure in the
MAS configuration are given. As mentioned in the previous section, we assume
$M=N$ so that in the MAS configuration we do not need to divide the received
symbols into different groups to implement the adjustable STC scheme. Take the
$2 \times 2$ Alamouti STBC scheme as an example, the adjustable STC scheme is
encoded as:
\begin{equation}\label{3.1.5}
\begin{aligned}
    {\boldsymbol C}_{rand} &= {\boldsymbol V}{\boldsymbol C}
    = \left[\begin{array}{cc} v_1 & 0 \\ 0 & v_2\end{array}\right]
    \left[\begin{array}{cc} r_{SR_k}1 & -r^*_{SR_k}2 \\ r_{SR_k}2 & ~r^*_{SR_k}1 \end{array} \right]\\ & = \left[\begin{array}{cc}v_1r_{SR_k}1 & -v_1r^*_{SR_k}2 \\ v_2r_{SR_k}2 & v_2r^*_{SR_k}1\end{array}\right],
\end{aligned}
\end{equation}
where $r_{SR_k}1$ and $r_{SR_k}2$ are the first symbols in the separate groups, and the $2 \times 2$ matrix ${\boldsymbol V}$ denotes the randomized matrix whose elements at the main diagonal are generated randomly according to different criteria described in \cite{B.Sirkeci-Mergen}. The transmission of the randomized STC schemes is described in (\ref{3.1.3}) and the decoding is given in (\ref{2.1.6.1}).

\section{Adaptive Buffer-Aided STC And Relay Optimization Algorithms}

In this section, the proposed ABARO algorithm in SAS is derived in detail. The
optimization in MAS follows a similar procedure with different channel vectors
so that we will skip the derivation. The main idea of the ABARO algorithm is to
choose the best relay node which contains the highest instantaneous $SNR$ for
transmission and reception in order to achieve full diversity order and higher
coding gain as compared to standard STC and DSTC designs. The relay nodes are
assumed to contain buffers to store the received data and forward the data to
the destination over the best available channels. In addition, the best relay
node is always chosen in order to enhance the detection performance at the
destination. As a result, with buffer-aided relays the proposed ABARO algorithm
will result in improved performance.

Before each transmission, the  {instantaneous $SNR$ ($SNR_{ins}$)}
of the $SR$ and $RD$ links are calculated at the destination and
conveyed with the help of signaling and feedback channels
\cite{A.Ikhlef}. The expressions for the instantaneous $SNR$ of the
$SR$ and $RD$ links are respectively given by {
\begin{equation}\label{2.1.7}
    SNR_{SR_k}[i] = \frac{\|f_{SR_k}[i]\|^2_F}{\sigma^2_r}, ~SNR_{R_kD}[i] =
\frac{\|{\boldsymbol V}_{eq}[i]g_{R_kD}[i]\|^2_F}{\sigma^2_d},
\end{equation}}
and the best link is chosen according to
\begin{equation}\label{2.1.8}
    SNR_{\rm opt}[i] = \arg\max_{k,b} SNR_{{\rm ins}_{k,b}}[i], ~k=1,2,...,n_r,~b=1,2,...,B, ~i=1,2,...,M/N,
\end{equation}
where $b$ denotes the occupied number of packets in the buffer.
After the best relay is determined, the transmission described in
(\ref{2.1.1}) and (\ref{2.1.2}) is implemented.  {The $SNR_{ins}$ is
calculated first and then the destination chooses a suitable relay
which has enough room in the buffer for the incoming data. For
example, if the $k$th $SR$ link is chosen but the buffer at the
$k$th relay node is full, the destination node will skip this node
and check the state of the buffer which has the second best link. In
this case the optimal relay with maximum instantaneous $SNR$ and
minimum buffer occupation at a certain $SNR$ level will be chosen
for transmission.}

After the detection of the first group of the received symbol vector at the
destination node, the adjustable code ${\boldsymbol v}$ will be optimized. The
constrained ML optimization problem that involves the detection of the
transmitted symbols and the computation of the adjustable code matrix at the
destination is written as
\begin{equation}\label{2.1.9}
\begin{aligned}
     & \left[\hat {\boldsymbol s}[i], \hat {\boldsymbol V}_{eq}[i]\right]  = \argmin_{{\boldsymbol s}[i],{\boldsymbol V}_{eq}[i]} \|{\boldsymbol r}[i] - \sqrt{P_RP_S}{\boldsymbol V}_{eq}[i]h[i]\hat{\boldsymbol s}[i]\|^2, \\
     & ~~~~~~~s.t.~ {\rm Tr}({\boldsymbol V}_{eq}[i]{\boldsymbol V}^\emph{H}_{eq}[i])\leq {P_{\boldsymbol V}}, ~i=1,2,...,M/N,
\end{aligned}
\end{equation}
where ${\boldsymbol r}[i]$ is the received symbol vector in the $i$th group and
$\hat{\boldsymbol s}[i]$ denotes the detected symbol vector in the $i$th group.
For example, if the number of antennas $N=4$ and the number of symbols stored
at the buffer is $M=8$, we have $M/N=2$ groups of symbols to implement the
adjustable STC scheme. According to the properties of the adjustable code
vector, the computation of $\hat{\boldsymbol s}[i]$ is the same as the decoding
procedure of the original STC schemes. In order to obtain the optimal code
vector ${\boldsymbol v}[i]$, the cost function in (\ref{2.1.9}) should be
minimized with respect to the equivalent code matrix ${\boldsymbol V}_{eq}[i]$
subject to a constraint on the transmitted power. The Lagrangian expression of
the optimization problem in (\ref{2.1.9}) is given by
\begin{equation}\label{2.1.10}
\begin{aligned}
    \mathcal {L} =& \|{\boldsymbol r}[i] - \sqrt{P_RP_S}{\boldsymbol V}_{eq}[i]h[i]\hat{\boldsymbol s}[i]\|^2 + \lambda(Tr({\bf V}_{\rm eq}[i] {\bf V}_{\rm eq}^\emph{H}[i]) - P_{\boldsymbol V}).
\end{aligned}
\end{equation}
It is worth mentioning that the power constraint expressed in (\ref{2.1.9}) is
ignored during the optimization of the adjustable code and in order to enforce
the power constraint, we introduce a normalization procedure after the
optimization which reduces the computational complexity. A stochastic gradient
algorithm is used to solve the optimization algorithm in (\ref{2.1.10}) with
lower computational complexity as compared to least-squares algorithms which
require the inversion of matrices. By taking the instantaneous gradient of
$\mathcal {L}$, discarding the power constraint and equating it to zero, we
obtain
\begin{equation}\label{2.1.11}
    \nabla\mathcal {L} = -\sqrt{P_RP_S}({\boldsymbol r}[i] - \sqrt{P_RP_S}{\boldsymbol V}_{eq}[i]h\hat{\boldsymbol s}[i])\hat{\boldsymbol s}^\emph{H}[i]h^\emph{H},
\end{equation}
and the ABARO algorithm for the proposed scheme can be expressed as follows
\begin{equation}\label{2.1.12}
\begin{aligned}
    {\boldsymbol V}_{eq}[i+1] = & {\boldsymbol V}_{eq}[i] - \mu\sqrt{P_RP_S}({\boldsymbol r}[i] - \sqrt{P_RP_S}{\boldsymbol V}_{eq}[i]h\hat{\boldsymbol s}[i])\hat{\boldsymbol s}^\emph{H}[i]h^\emph{H}[i],
\end{aligned}
\end{equation}
where $\mu$ is the step size.  {After the update of the equivalent
coding matrix ${\boldsymbol V}_{eq}$ in SAS, we can recover the
original coding vector ${\boldsymbol v}[i]$ from the entries of the
main diagonal of ${\boldsymbol V}_{eq}$. A normalization of the
original code vector ${\boldsymbol v}[i]$ that circumvents the power
constraint in (\ref{2.1.9}) is given by
\begin{equation}\label{2.1.13}
    {\boldsymbol v}[i+1] = {\boldsymbol v}[i+1]\frac{P_{\boldsymbol v}}{\sqrt{{\boldsymbol v}^\emph{H}[i+1]{\boldsymbol v}[i+1]}}.
\end{equation}}
 {Similarly, the ABARO algorithm in the MAS configuration can be
implemented step-by-step as shown in (\ref{2.1.7}) to (\ref{2.1.13}).} A
summary of the ABARO algorithm in the MAS configuration is shown in Table I.

\begin{table*}
  \centering
  \caption{Summary of the Adaptive Buffer-Aided Relaying Optimization Algorithm for MAS configuration}\label{t1}
  \begin{tabular}{|l|}
  \hline
  \bfseries{Initialization:}\\
  \qquad Empty the buffer at the relays,\\
  \bfseries{for $j=1,2,...$} \\

  \qquad \bfseries{if $j=1$} \\
  \qquad\qquad compute: $ {SNR_{SR_k}[j] = \frac{\|\boldsymbol{F}_{SR_k}[j]\|_F^2}{\sigma^2_n}, ~k=1,2,...,n_r,}$\\
  \qquad\qquad compare: $SNR_{opt}[j] = \arg\min_{k,b} SNR^{-1}_{{\rm ins}_{k,b}}[j], ~k=1,2,~...~,n_r,~b=1,2,~...~,B,$\\
  \qquad\qquad $\boldsymbol{r}_{SR_k}[j] = \sqrt{\frac{P_S}{N}}\boldsymbol{F}_{SR_k}[j]\boldsymbol{s}[j] + \boldsymbol{n}_{SR_k}[j]$,\\

  \qquad \bfseries{else}\\
  \qquad\qquad compute: $ {SNR_{SR_k}[j] = \frac{\|\boldsymbol{F}_{SR_k}[j]\|_F^2}{\sigma^2_n}, ~k=1,2,...,n_r}$\\
  \qquad\qquad\qquad\qquad $ {SNR_{R_kD}[j] = \frac{\|{\boldsymbol V}_{eq}[j]{\boldsymbol G}_{R_kD}[j]\|^2_F}{\sigma^2_d}, ~k=1,2,...,n_r}$,\\
  \qquad\qquad compare: $SNR_{opt}[j] = \arg\max \{SNR_{SR_k}[j],SNR_{R_kD}[j]\}$, ~$k=1,2, ... ,n_r$,\\
  \qquad\qquad\qquad \bfseries{if} $SNR_{max}[j] = SNR_{SR_k}[j] ~\& \rm{~Relay_k ~is ~not ~full}$\\
  \qquad\qquad\qquad\qquad $\boldsymbol{r}_{SR_k}[j] = \sqrt{\frac{P_S}{N}}\boldsymbol{F}_{SR_k}[j]\boldsymbol{s}[j] + \boldsymbol{n}_{SR_k}[j]$,\\
  \qquad\qquad\qquad \bfseries{elseif} $SNR_{max}[j] = \arg\max SNR_{R_kD}[j] ~\&  \rm{ ~Relay_k ~is ~not ~empty}$\\
  \qquad\qquad\qquad\qquad ${\boldsymbol r}_{R_kD}[j] = \sqrt{\frac{P_R}{N}}{\boldsymbol V}_{eq}[j]{\boldsymbol H}[j]{\boldsymbol s}[j] + {\boldsymbol n}[j]$,\\
  \qquad\qquad\qquad\qquad ML detection: \\
  \qquad\qquad\qquad\qquad\qquad $\hat{\boldsymbol s}[j] = \arg\min_{\hat{\boldsymbol s}[j]} \|{\boldsymbol r}_{R_kD}[j] - \sqrt{\frac{P_RP_S}{N}}{\boldsymbol V}_{eq}[j]{\boldsymbol H}[j]\hat{\boldsymbol s}[j]\|^2$,\\
  \qquad\qquad\qquad\qquad Adjustable Matrix Optimization: \\
  \qquad\qquad\qquad\qquad\qquad ${\boldsymbol V}_{eq}[j+1] = {\boldsymbol V}_{eq}[j] - \mu\sqrt{\frac{P_R P_S}{N}}({\boldsymbol r}_{R_kD}[j] - \sqrt{\frac{P_RP_S}{N}}{\boldsymbol V}_{eq}[j]{\boldsymbol H}[j]\hat{\boldsymbol s}[j])\hat{\boldsymbol s}^\emph{H}[j]{\boldsymbol H}^\emph{H}[j]$,\\
  \qquad\qquad\qquad\qquad Normalization: \\
  \qquad\qquad\qquad\qquad\qquad ${\boldsymbol V}[j+1] = {\boldsymbol V}[j+1]\frac{P_{\boldsymbol V}}{\sqrt{\|{\boldsymbol V}[j+1]\|_F^2}}$,\\
  \qquad\qquad\qquad \bfseries{elseif} $SNR_{SR_k} \rm{~is ~max} ~\&  \rm{ ~Relay_k ~is ~full}$\\
  \qquad\qquad\qquad\qquad skip this Relay,\\
  \qquad\qquad\qquad \bfseries{elseif} $SNR_{R_kD} \rm{~is ~max} ~\&  \rm{ ~Relay_k ~is ~empty}$\\
  \qquad\qquad\qquad\qquad skip this Relay,\\
  \qquad\qquad\qquad\qquad\qquad ...repeat...\\
  \qquad\qquad\qquad \bfseries{end}\\
  \qquad \bfseries{end}\\
  \bfseries{end}\\

\hline
\end{tabular}
\end{table*}

\section{Best Relay Selection with DSTC Schemes}

In this section, we assume that the relays contain buffers and employ DSTC
schemes in the second hop for the SAS and MAS configurations. In particular, we
also present the design of a best group relay selection algorithm for
performance enhancement. The details of the deployment of DSTC schemes in the
MAS configuration is similar to that in the SAS scheme. Therefore, we will not
repeat it to avoid redundancy. The main difference between the relay selection
algorithm for DSTC schemes as compared to that for STC schemes is due to the
fact that for DSTC schemes a group of relays is selected. Specifically for DSTC
schemes, the source node broadcasts data to all the relays and a DF protocol is
employed at the relays. After the detection, the proposed group relay selection
algorithm is employed. It is important to notice that if the DSTC schemes are
used at the relays, each relay has to contain one copy of the modulated symbol
vector which means in the first hop the source node cannot choose the best
relay but only broadcast the symbol vector to all relays. The adjustable code
vectors can be considered at each relay as well.

\subsection{DSTBC schemes}

In this subsection, we detail the DSTBC scheme used in this study. In the SAS
configuration, a single antenna is used in each node and the DF protocol is
employed at the relay nodes. In the first hop, the source node broadcasts
information symbol vector ${\boldsymbol s}$ to the relay node which is given by
\begin{equation}\label{4.0.1}
\begin{aligned}
    \boldsymbol{r}_{SR_k}[j] &= \sqrt{P_S}f_{SR_k}[j]\boldsymbol{s}[j] + \boldsymbol{n}_{SR_k}[j], ~k=1,2,...,n_r, ~j = 1,2,...,J,
\end{aligned}
\end{equation}
where $\boldsymbol{s}[j]$ is a block of symbols with length of $M$,
$f_{SR_k}[j]$ denotes the CSI and $\boldsymbol{n}_{SR_k}[j]$ stands
for the $M \times 1$ AWGN. The transmission power assigned at the
source node is denoted as $P_S$. After the detection at the $k$th
node, $\hat{\boldsymbol{s}}_k$ can be obtained. The relays are then
divided into $m=N_{DSTC}/n_r$ groups to implement the DSTC scheme,
where $N_{DSTC}$ denotes the number of antennas to form the DSTC
scheme. It should be noted that synchronization at the symbol level
and of the carrier phase is assumed in this work.  {If one considers
the distributed Alamouti STBC as an example, the encoding procedure
is detailed in Table II, where ${\boldsymbol s} = [s^{(1)}_1$
$s^{(1)}_2]$ denotes the estimated symbols at relay $1$, and
${\boldsymbol s} = [s^{(2)}_1 s^{(2)}_2]$ denotes the symbols
estimated at relay $2$.}
\begin{table*}
  \centering
  \caption{Distributed Alamouti in SAS}\label{t1}
  \begin{tabular}{|c|c|c|}
  \hline
  ~ & 1st Time Slot & 2nd Time Slot \\ \hline
  Relay 1 & $s^{(1)}_1$ & $-s^{(1)*}_2$ \\ \hline
  Relay 2 & $s^{(2)}_2$ & $-s^{(2)*}_1$ \\
\hline
\end{tabular}
\end{table*}
Note that it is assumed that the best relays will be chosen in the second hop
and synchronization is perfect so after the relays forward the DSTC schemes to
the destination, a composite signal comprising DSTC transmissions from multiple
relays is received. The signal received in the second hop is described by
\begin{equation}\label{4.0.2}
\begin{aligned}
    \boldsymbol{r}_{RD_m}[j] &= \sum_{m=1}^{N_{DSTC}/n_r}\sqrt{\frac{P_R}{N_{DSTC}}}{\boldsymbol g}_{RD_m}[j]{\boldsymbol C}_m[j] + \boldsymbol{n}_{RD_m}[j], ~j=1,2,...,M/N_{DSTC}, ~m = 1,2,...,N_{DSTC}/n_r,
\end{aligned}
\end{equation}
where $\boldsymbol{r}_{RD_m}[j]$ denotes the $T \times 1$ received symbol
matrix, and ${\boldsymbol g}_{RD_m}[j]$ denotes the $m$th channel coefficients
vector. The parameter $M$ denotes the number of symbols stored in the buffers,
$m$ denotes the number of relay groups to implement the DSTC scheme and $j$
denotes the DSTC scheme index.

\subsection{Best Relay Selection with DSTC in SAS}

In this subsection, we describe the best relay selection algorithm used in
conjunction with the DSTC scheme in the SAS configuration. In particular, the
best relay selection algorithm is based on the techniques reported in
\cite{A.Bletsas} and \cite{N.Nomikos}, however, the approach presented here is
modified for DSTC schemes and buffer-aided relay systems. In the first hop, the
$M \times 1$ modulated signal vector ${\boldsymbol s}[j]$ is broadcast to the
relays during $M$ time slots and the $M \times 1$ received symbol vector
${\boldsymbol r}_{SR_k}[j]$ is given by
\begin{equation}\label{4.1}
{\boldsymbol r}_{SR_k}[j] = \sqrt{P}f_{SR_k}[j]{\boldsymbol s}[j] +
{\boldsymbol n}[j], ~k = 1,2,...,n_r, ~j = 1,2,...,J,
\end{equation}
where $f_{SR_k}[j]$ denotes the complex scalar channel gain between
the $k$th relay and the destination, and the AWGN noise vector
${\boldsymbol n}[j]$ is generated at the $k$th relay node with
variance equal to $\sigma^2_n$. The relays are equipped with buffers
to store the received symbol vectors and the optimal relays are
chosen according to the approach reported in \cite{B.Maham} in order
to implement the DSTC scheme among the relays. Specifically, all the
relays will be divided into $m=\frac{N_{DSTC}}{n_r}$ groups and the
best relay group with the highest $SINR$ will be chosen to forward
the received symbols. The opportunistic relay selection algorithm is
given by {
\begin{equation}\label{4.2}
SINR_k[j] = \argmax_{{\boldsymbol g}_{RD_k}[j]} \frac{\|{\boldsymbol
g}_{RD_k}[j]\|^2_F}{\sum_{m=1,m\neq k}^{K}\sqrt{\|{\boldsymbol
g}_{RD_m}[j]\|^2_F} + \sigma^2_d},
\end{equation}}
where ${\boldsymbol g}_{RD_m}[j]$ denote the $1 \times N_{DSTC}$ channel vector
between the chosen relays and the destination to implement the DSTC scheme and
$K=C_{n_r}^{N_{DSTC}}$ denotes all possible relay group combinations. The noise
variance is given by $\sigma^2_d$. After the relay group selection, the optimal
relay group transmits the DSTC signals to the destination node and the received
data at the destination is described by
\begin{equation}\label{4.3}
\begin{aligned}
    \boldsymbol{r}_{RD_m}[j] &=
    \sqrt{\frac{P_R}{N_{DSTC}}}{\boldsymbol g}_{RD_m}[j]{\boldsymbol C}_m[j] + \boldsymbol{n}_{RD_m}[j],
\end{aligned}
\end{equation}
where ${\boldsymbol C}_m[j]$ denotes the DSTC scheme encoded among the chosen
relays. The DSTC decoding process is similar to that of the original STC
scheme. It is worth mentioning that the adjustable coding schemes can be
introduced in DSTC schemes and the optimization of the adjustable code vector
will result in a performance improvement. The summary of the ABARO algorithm
for DSTC schemes in the SAS configuration is shown in Table III.

\begin{table*}
  \centering
  \caption{Summary of the Adaptive Buffer-Aided Relaying Optimization Algorithm for DSTC Schemes in SAS}\label{t1}
  \begin{tabular}{|l|}
  \hline
  \bfseries{Initialization:}\\
  \qquad Empty the buffer at the relays,\\
  \bfseries{for $j=1,2,...$} \\
  \qquad \bfseries{if $j=1$} \\
  \qquad\qquad ${\boldsymbol r}_{SR_k}[j] = \sqrt{P_S}f_{SR_k}[j]{\boldsymbol s}[j] + {\boldsymbol n}[j]$,\\
  \qquad \bfseries{else}\\
  \qquad\qquad compute: $ {SNR_{SR_k}[j] = \sum_{k=1}^{n_r}\frac{\|f_{SR_k}[j]\|_F^2}{\sigma^2_n}}$,\\
  \qquad\qquad\qquad\qquad $ {SNR_{R_kD}[j] = \sum_{k=1}^{n_r}\frac{\|g_{R_kD}[j]\|^2_F}{\sigma^2_d}}$,\\
  \qquad\qquad compare: $SNR_{opt}[j] = \arg\max \{SNR_{SR_k}[j],SNR_{R_kD}[j]\}$,\\
  \qquad\qquad\qquad \bfseries{if} $SNR_{max}[j] = SNR_{SR_k}[j] ~\&  \rm{~All ~the ~Relays ~are ~not ~full}$\\
  \qquad\qquad\qquad\qquad $\boldsymbol{r}_{SR_k}[j] = \sqrt{P_S}f_{SR_k}[j]\boldsymbol{s}[j] + \boldsymbol{n}_{SR_k}[j]$,\\
  \qquad\qquad\qquad \bfseries{elseif} $SNR_{max}[j] = SNR_{R_kD}[j] ~\& ~All ~the ~Relays \rm{ ~are ~not ~empty}$\\
  \qquad\qquad\qquad\qquad\qquad $ {SINR_k[j] = \arg\max_{{\boldsymbol g}_{RD_k}[j]} \frac{\|{\boldsymbol g}_{RD_m}[j]\|^2_F}{\sum_{m=1,m\neq k}^{K}\sqrt{\|{\boldsymbol g}_{RD_m}[j]\|^2_F} + \sigma^2_d}}$,\\
  \qquad\qquad\qquad\qquad\qquad $\boldsymbol{r}_{R_kD}[j] = \sqrt{\frac{P_R}{N_{DSTC}}}{\boldsymbol g}_{R_kD}[j]{\boldsymbol C}_k[j] + \boldsymbol{n}_{R_kD}[j]$,\\
  \qquad\qquad\qquad \bfseries{elseif} $SNR_{SR_k}[j] \rm{~is ~max} ~\& ~Relay_k \rm{ ~is ~full}$\\
  \qquad\qquad\qquad\qquad skip this Relay,\\
  \qquad\qquad\qquad \bfseries{elseif} $SNR_{R_kD} \rm{~is ~max} ~\& ~Relay_k \rm{ ~is ~empty}$\\
  \qquad\qquad\qquad\qquad skip this Relay,\\
  \qquad\qquad\qquad\qquad\qquad ...repeat...\\
  \qquad\qquad\qquad \bfseries{end}\\
  \qquad \bfseries{end}\\
  \bfseries{end}\\

\hline
\end{tabular}
\end{table*}

\subsection{Best Relay Selection with DSTC in MAS}

The best relay selection algorithm described in the previous section is now
extended to the MAS configuration in this subsection. The main difference
between the best relay selection for SAS and MAS is the use of multiple
antennas at each node. Moreover, the relays equipped with multiple antennas
will obtain a complete STC scheme and only one best relay node will be chosen
according to the BRS algorithm. Assuming $M=N$, each node equips $N \geq 2$
antennas and in the first hop, the $M \times 1$ modulated signal vector
${\boldsymbol s}[j]$ is broadcast to the relays within $1$ time slot and the $M
\times 1$ received symbol matrix ${\boldsymbol r}_{SR_k}[j]$ is given by
\begin{equation}\label{4.4}
{\boldsymbol r}_{SR_k}[j] = \sqrt{\frac{P}{N}}{\boldsymbol
F}_{SR_k}[j]{\boldsymbol s}[j] + {\boldsymbol n}[j], ~k=1,2,...,n_r, ~j =
1,2,...J,
\end{equation}
where ${\boldsymbol F}_{SR_k}[j]$ denotes the channel coefficient
matrix between the $k$th relay and the destination, and the AWGN
noise vector ${\boldsymbol n}[j]$ is generated at the $k$th relay
node with variance $\sigma^2_n$. The $N \times 1$ received symbol
vector is stored at the relays and the optimal relay will be chosen
according to \cite{B.Maham}. The opportunistic relay selection
algorithm for the DSTC scheme and the MAS configuration is given by
{
\begin{equation}\label{4.5}
SNR_k[j] = \argmax_{{\boldsymbol G}_{R_kD}[j]} \frac{\|{\boldsymbol
G}_{R_kD}[j]\|^2_F}{\sigma^2_d}, ~k = 1,2,...,n_r,
\end{equation}}
where ${\boldsymbol G}_{R_kD}[j]$ denotes the $N \times N$ channel matrix
between the $k$th relay and the destination. After the best relay with the
maximum $SNR$ is chosen, the data is encoded by the DSTC scheme. The DSTC
encoded and transmitted data in the second hop is received at the destination
as described by
\begin{equation}\label{4.6}
{\boldsymbol R}[j] = \sqrt{\frac{P}{N}} {\boldsymbol G}_{R_kD}[j]{\boldsymbol
M}[j] + {\boldsymbol N}[j],
\end{equation}
where ${\boldsymbol M}[j]$ denotes the $N \times T$ DSTC encoded data,
${\boldsymbol R}[j]$ denotes the $N \times T$ received data matrix, and
${\boldsymbol N}[j]$ is the AWGN matrix with variance $\sigma^2_d$.

\section{Analysis}

 {In this section, we assess the computational complexity of the
proposed algorithms and derive the pairwise error probability (PEP) of
cooperative systems that employ adaptive STC and DSTC schemes.} The expression
of the PEP upper bound is adopted due to its relevance to assess STC and DSTC
schemes. We also study the effects of the use of buffers and adjustable codes
at the relays, and derive analytical expressions for their impact on the PEP.
As mentioned in Section II, the adjustable codes are considered in the
derivation as it affects the performance by reducing the upper bound of the
PEP. Similarly, the buffers store the data and forward it by selecting the best
available associated channel for transmission so that the performance
improvement is quantified in our analysis. The PEP upper bound of the
traditional STC schemes in \cite{Hamid} is used for comparison purposes. The
main difference between the PEP upper bound in \cite{Hamid} and that derived in
this section lies in the increase of the eigenvalues of the adjustable codes
and channels which leads to higher coding gains. The derived upper bound holds
for systems with different sizes and an arbitrary number of relay nodes.

 {
\subsection{Computational Complexity Analysis}
According to the description of the proposed algorithms in Sections IV and V,
the SG algorithms reduces the computational complexity by avoiding the channel
inversion as compared to the existing algorithms. The computational complexity
of the proposed SG adjustable matrix optimization in the SAS and MAS
configurations is $(3+T)N$ and $(3+T)N^2$, respectively. The main difference
between the proposed algorithms in the SAS and MAS configurations is the number
of antennas. For example, the computational complexity of $SNR$ in $SR$ and
$RD$ links in SAS configuration is $2N(1+T)$ according to (\ref{2.1.7}), while
the computational complexity of $SNR$ in $SR$ and $RD$ links in the MAS
configuration is $2N^2(1+T)$. In addition, if a higher-level modulation scheme
is employed, larger relay networks and more antennas are used at the relay
node, the STC and DSTC schemes and the relay selection algorithm as well as the
coding vector optimization algorithm become more complex. For example, if a
$4$-antenna relay node is employed, the number of multiplications will be
increased from $10$ when using a $2$-antenna relay node to $28$, and if $4$
single-antenna relay nodes are employed to implement a DSTC scheme the number
of multiplications will be increased from $20$ to $112$.}

\subsection{Pairwise Error Probability}

Consider an $N \times N$ STC scheme at the relay node with $T$ codewords. The
codeword ${\boldsymbol C}^1$ is transmitted and decoded as another codeword
${\boldsymbol C}^i$ at the destination node, where $i=1,2,...,T$. According to
\cite{Hamid}, the probability of error for this code can be upper bounded by
the sum of all the probabilities of incorrect decoding, which is given by
\begin{equation}\label{5.1.1}
    {\rm P_e} \leq \sum_{i=2}^{T} {\rm P}({\boldsymbol C}^1\rightarrow{\boldsymbol C}^i).
\end{equation}
Assuming that the codeword ${\boldsymbol C}^2$ is decoded at the destination
node and that we know the channel information perfectly, we can derive the
conditional PEP of the STC encoded with the adjustable code matrix
${\boldsymbol V}$ as \cite{JYuan}
\begin{equation}\label{5.1}
    {\rm P}({\boldsymbol C}^1\rightarrow{\boldsymbol C}^2\mid{{\boldsymbol V}})
    = {\rm Q}\left(\sqrt{\frac{\gamma}{2}}\parallel{\boldsymbol V}{\boldsymbol G}_{R_kD}({\boldsymbol C}^1-{\boldsymbol C}^2)\parallel_F\right),
\end{equation}
where ${\boldsymbol G}_{R_kD}$ stands for the channel coefficients matrix. Let
${\boldsymbol U}^\emph{H}{\boldsymbol \Lambda}_{\boldsymbol C}{\boldsymbol U}$
be the eigenvalue decomposition of $({\boldsymbol C}^1-{\boldsymbol
C}^2)^\emph{H}({\boldsymbol C}^1-{\boldsymbol C}^2)$, where ${\boldsymbol U}$
is a unitary matrix with the eigenvectors and ${\boldsymbol
\Lambda}_{\boldsymbol C}$ is a diagonal matrix which contains all the
eigenvalues of the difference between two different codewords ${\boldsymbol
C}^1$ and ${\boldsymbol C}^2$. Let ${\boldsymbol Y}^\emph{H}{\boldsymbol
\Lambda}_{{\boldsymbol G}_n}{\boldsymbol Y}$ stand for the eigenvalue
decomposition of $({\boldsymbol G}_{RkD}{\boldsymbol U})^H{\boldsymbol
G}_{RkD}{\boldsymbol U}$, where ${\boldsymbol Y}$ is a unitary matrix that
contains the eigenvectors and ${\boldsymbol \Lambda}_{\boldsymbol V}$ is a
diagonal matrix with the eigenvalues arranged in decreasing order. The
eigenvalue decomposition of $({\boldsymbol Y}{\boldsymbol V}{\boldsymbol
U})^\emph{H}{\boldsymbol Y}{\boldsymbol V}{\boldsymbol U}$ is denoted by
${\boldsymbol W}^\emph{H}{\boldsymbol \Lambda}_{{\boldsymbol V}_n}{\boldsymbol
W}$, where ${\boldsymbol W}$ is a unitary matrix that contains the eigenvectors
and ${\boldsymbol \Lambda}_{{\boldsymbol V}_n}$ is a diagonal matrix with the
eigenvalues. Therefore, the conditional PEP can be written as
\begin{equation}\label{5.2}
    {\rm P}({\boldsymbol C}^1\rightarrow{\boldsymbol C}^2\mid{\boldsymbol V})={\rm Q}\left(\sqrt{\frac{\gamma}{2}\sum^{NT}_{m=1}\sum^N_{n=1}\lambda_{{\boldsymbol V}_n}\lambda^{opt}_{{\boldsymbol G}_n}\lambda_{{\boldsymbol C}_n}|\xi_{n,m}|^2}\right),
\end{equation}
where $\xi_{n,m}$ is the $(n,m)$th element in ${\boldsymbol Y}$, and
$\lambda_{{\boldsymbol V}_n}$, $\lambda^{opt}_{{\boldsymbol G}_n}$ and
$\lambda_{{\boldsymbol C}_n}$ are the $n$th eigenvalues in ${\boldsymbol
\Lambda}_{\boldsymbol V}$, ${\boldsymbol \Lambda}_{{\boldsymbol G}_n}$ and
${\boldsymbol \Lambda}_{\boldsymbol C}$, respectively. It is important to note
that the value of $\lambda_{\boldsymbol V}$ and $\lambda^{opt}_{{\boldsymbol
G}_n}$ are positive and real because $({\boldsymbol G}_{R_kD}{\boldsymbol
U})^H{\boldsymbol G}_{RkD}{\boldsymbol U}$ and $({\boldsymbol Y}{\boldsymbol
V}{\boldsymbol U})^\emph{H}{\boldsymbol Y}{\boldsymbol V}{\boldsymbol U}$ are
Hermitian symmetric matrices. According to \cite{Hamid}, an appropriate upper
bound assumption of the ${\rm Q}$ function is ${\rm
Q}(x)\leq\frac{1}{2}e^{\frac{-x^2}{2}}$, thus the upper bound of the PEP for an
adaptive STC scheme is given by
\begin{equation}\label{5.3}
    {\rm P}_{e_{\boldsymbol V}}\leq E \left[\frac{1}{2}\exp\left(-\frac{\gamma}{4}\sum^{NT}_{m=1}\sum^N_{n=1}\lambda_{{\boldsymbol V}_n}\lambda^{opt}_{{\boldsymbol G}_n}\lambda_{{\boldsymbol C}_n}|\xi_{n,m}|^2\right) \right]=\frac{1}{\prod_{n=1}^{N}(1+\frac{\gamma}{4}\lambda_{{\boldsymbol V}_n}\lambda^{opt}_{{\boldsymbol G}_n}\lambda_{{\boldsymbol C}_n})^{NT}}.
\end{equation}
The key elements of the PEP are $\lambda_{{\boldsymbol V}_n}$ and
$\lambda^{opt}_{{\boldsymbol G}_n}$ which related to the adjustable code
matrices and the channels in the second hop. In the following subsection we
will provide an analysis of these key elements separately.

\subsection{Effect of Adjustable Code Matrices}

Before the analysis of the effect of the adjustable code matrices, we derive
the expression of the upper bound of the error probability expression for a
traditional STC. It is worth mentioning that in this section, we focus on the
effort of using adjustable code matrices at the relays and the relay selection
and the effort of buffers are not considered.

According to \cite{Hamid}, the PEP upper bound of the SAS configuration using
traditional STC schemes is given by
\begin{equation}\label{5.4}
    {\rm P}_{e} \leq E \left[\frac{1}{2}\exp\left(-\frac{\gamma}{4}\sum^{NT}_{m=1}\sum^N_{n=1}\lambda_{{\boldsymbol C}_n}|\xi_{n,m}|^2\right)\right]=\frac{1}{\prod_{n=1}^{N}(1+
    \frac{\gamma}{4}\lambda_{{\boldsymbol C}_n})^{NT}},
\end{equation}
where $\lambda_{{\boldsymbol C}_n}$ denotes the $n$th eigenvalue of the
distance matrix by using a traditional STC scheme. If we rearrange the terms in
(\ref{5.4}), we can rewrite the upper bound of the PEP of traditional STC
scheme as
\begin{equation}\label{5.5}
    {\rm P}_{e} \leq \left(\frac{\gamma}{4}\right)^{-N^2T}\prod_{n=1}^{N}\lambda^{-NT}_{{\boldsymbol C}_n}.
\end{equation}
If we only consider adjustable code matrices at relays without the relay
selection and buffers, the upper bound of the PEP of the proposed ABARO
algorithm is derived as
\begin{equation}\label{5.6}
    {\rm P}_{e_{\boldsymbol V}}\leq \frac{1}{\prod_{n=1}^{N}(1+\frac{\gamma}{4}\lambda_{{\boldsymbol V}_n}\lambda_{{\boldsymbol C}_n})^{NT}} \approx \left(\left(\frac{\gamma}{4}\right)^{-N^2T}\prod_{n=1}^{N}\lambda^{-NT}_{{\boldsymbol C}_n}\right)\prod_{n=1}^{N}\lambda^{-NT}_{{\boldsymbol V}_n} = {\rm P}_{e}\prod_{n=1}^{N}\lambda^{-NT}_{{\boldsymbol V}_n},
\end{equation}
By comparing (\ref{5.5}) and (\ref{5.6}), employing an adjustable code matrix
for an STC scheme at the relay node introduces $\lambda_{{\boldsymbol V}_n}$ in
the PEP upper bound. The adjustable code matrices are chosen according to the
criterion introduced in \cite{B.Sirkeci-Mergen} and the Hermitian matrix
${\boldsymbol V}^\emph{H}_n{\boldsymbol V}_n$ is positive semi-definite. With
the aid of numerical tools, we have found that ${\boldsymbol
\Lambda}_{\boldsymbol V}$ is diagonal with one eigenvalue less than $1$ and
others much greater than $1$. We define the coding gain factor $\eta$ which
denotes the quotient of the traditional STC PEP and the adjustable STC PEP as
described by
\begin{equation}\label{5.6.1}
    \eta \triangleq \frac{P_e}{P_{e_{\boldsymbol V}}} = \prod_{n=1}^{N}\lambda^{NT}_{{\boldsymbol V}_n} \gg 1.
\end{equation}
As a result, by using the adjustable code matrices at the relays contributes to
a decrease of the BER performance. The effect of employing and optimizing the
adjustable code matrix corresponds to introducing coding gain into the STC
schemes. The power constraint enforced by (\ref{2.1.13}) introduces no
additional power and energy during the optimization. As a result, employing the
adjustable code matrices in the MAS and the SAS configurations can provide a
decrease in the BER upper bound since the value in the denominator increases
without additional transmit power.

\subsection{Effect of Buffer-aided Relays}

In this subsection, the effect of using buffers at the relays is mathematically
analyzed. The expression of the PEP upper bound is adopted again in this
subsection. The traditional STC scheme is employed in this subsection in order
to highlight the performance improvement by using buffers at the relays.

Let ${\boldsymbol U}^\emph{H}{\boldsymbol \Lambda}_{\boldsymbol C}{\boldsymbol
U}$ be the eigenvalue decomposition of $({\boldsymbol C}^1-{\boldsymbol
C}^2)^\emph{H}({\boldsymbol C}^1-{\boldsymbol C}^2)$ and ${\boldsymbol
Y}^\emph{H}{\boldsymbol \Lambda}_{{\boldsymbol G}_{R_kD}}{\boldsymbol Y}$ be
the eigenvalue decomposition of $({\boldsymbol G}_{RkD}{\boldsymbol
U})^H{\boldsymbol G}_{RkD}{\boldsymbol U}$, the PEP upper bound of a
traditional STC scheme in buffer-aided relays is given by
\begin{equation}\label{5.7}
\begin{aligned}
    P_{e_{{\boldsymbol G}^{opt}_n}} & \leq E \left[\frac{1}{2}\exp\left(-\frac{\gamma}{4}\sum^{NT}_{m=1}\sum^N_{n=1}\lambda^{opt}_{{\boldsymbol G}_n}\lambda_{{\boldsymbol C}_n}|\xi_{n,m}|^2\right)\right]=\frac{1}{\prod_{n=1}^{N}(1+
    \frac{\gamma}{4}\lambda^{opt}_{{\boldsymbol G}_n}\lambda_{{\boldsymbol C}_n})^{NT}}\\ & \approx \left(\frac{\gamma}{4}\right)^{-N^2T}\prod_{n=1}^{N}\lambda^{-NT}_{{\boldsymbol C}_n}\prod_{n=1}^{N}\lambda^{-NT}_{{\boldsymbol G}^{opt}_n},
\end{aligned}
\end{equation}
where $\lambda_{{\boldsymbol C}_n}$ denotes the eigenvalues of the traditional
STC scheme and $\lambda_{{\boldsymbol G}^{opt}_n}$ denotes the eigenvalue of
the channel components. The PEP performance of a traditional STC scheme without
buffer-aided relays is given by
\begin{equation}\label{5.8}
    P_e \leq \frac{1}{\prod_{n=1}^{N}(1+
    \frac{\gamma}{4}\lambda_{{\boldsymbol G}_n}\lambda_{{\boldsymbol C}_n})^{NT}} \approx \left(\frac{\gamma}{4}\right)^{-N^2T}\prod_{n=1}^{N}\lambda^{-NT}_{{\boldsymbol C}_n}\prod_{n=1}^{N}\lambda^{-NT}_{{\boldsymbol G}_n},
\end{equation}
where $\lambda_{{\boldsymbol C}_n}$ denotes the eigenvalues of the traditional
STC scheme and $\lambda_{{\boldsymbol G}_n}$ denotes the eigenvalue of the
channels in second hop. By comparing (\ref{5.7}) and (\ref{5.8}), the only
difference is the product of the channel eigenvalues. To show the advantage of
employing buffer-aided relays, we need to prove that $P_{e_{{\boldsymbol
G}^{opt}_n}} < P_e$.

We can simply divide (\ref{5.7}) by (\ref{5.8}) and obtain
\begin{equation}\label{5.9}
    \beta = \frac{P_e}{P_{e_{{\boldsymbol G}^{opt}_n}}} =
    \frac{\left(\frac{\gamma}{4}\right)^{-N^2T}\prod_{n=1}^{N}\lambda^{-NT}_{{\boldsymbol C}_n}\prod_{n=1}^{N}\lambda^{-NT}_{{\boldsymbol G}_n}}{\left(\frac{\gamma}{4}\right)^{-N^2T}\prod_{n=1}^{N}\lambda^{-NT}_{{\boldsymbol C}_n}\prod_{n=1}^{N}\lambda^{-NT}_{{\boldsymbol G}^{opt}_n}} = \frac{\prod_{n=1}^{N}\lambda^{NT}_{{\boldsymbol G}^{opt}_n}}{\prod_{n=1}^{N}\lambda^{NT}_{{\boldsymbol G}_n}}.
\end{equation}
As derived in Section IV, the instantaneous SNR of the channels is
computed and the channel with highest SNR is chosen which contains
the largest eigenvalues among all the channels.  {As a result, we
have
\begin{equation}\label{5.9.1}
    \lambda^{opt}_{{\boldsymbol C}_n} > \lambda_{{\boldsymbol C}_n}, n = 1,2,...,N,
\end{equation}
which gives
\begin{equation}\label{5.10}
    \beta = \frac{P_e}{P_{e_{{\boldsymbol G}^{opt}_n}}} = \frac{\prod_{n=1}^{N}\lambda^{NT}_{{\boldsymbol G}^{opt}_n}}{\prod_{n=1}^{N}\lambda^{NT}_{{\boldsymbol G}_n}} \gg 1.
\end{equation}
Through (\ref{5.10}), we have proved that $P_{e_{{\boldsymbol G}^{opt}_n}} <
P_e$ which indicates the BER performance of a system that employs buffer-aided
relays is improved as compared to that of a system using relays without
buffers.}

\section{Simulation}

The simulation results are provided in this section to assess the
proposed scheme and algorithms in the SAS and the MAS
configurations. In this work, we consider the AF protocol with the
standard Alamouti STBC scheme and randomized Alamouti (R-Alamouti)
scheme in \cite{B.Sirkeci-Mergen}. The BPSK modulation is employed
and each link between the nodes is characterized by static block
fading with AWGN.  {The period during which the channel is static is
equal to one symbol transmission period in Figs. 4, 5 and 6, whereas
in Figs. 3 and 7 such period is equal to one packet size. The packet
size is $M=100$ symbols and the number of packets is $J=200$. The
effects of different buffer sizes are also evaluated.} Different STC
schemes can be employed with a simple modification as well as the
proposed relay selection and ABARO algorithms can be incorporated.
We employ $n_r=1,2$ relay nodes and $N=1,2$ antennas at each node,
and we set the symbol power $\sigma^2_s$ to 1.

\begin{figure}
\begin{center}
\def\epsfsize#1#2{0.65\columnwidth}
\epsfbox{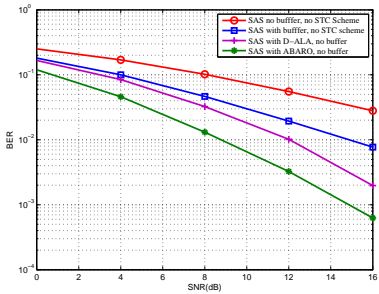}\vspace*{-1em} \caption{Buffer v.s. No Buffer in SAS}\label{1}
\vspace{-1em}
\end{center}
\end{figure}

The upper bounds of the D-Alamouti, the proposed ABARO algorithm and the
buffer-aided relays in the SAS configurations are shown in Fig. 3.
 {The theoretical PEP result of a standard SAS configuration,
which does not employ STC schemes or buffer-aided relays, is shown as the curve
contains the largest decoding errors. By comparing the first two BER curves in
Fig. 3 we can conclude that by employing buffers at relays, the decoding error
upper bound is decreased. In this case, the effect of using buffers at the
relays contributes to reducing the PEP performance dramatically. If the STC
scheme is employed at the relays, an increase of diversity order is observed in
Fig. 3. By comparing the lower BER curves in Fig. 3, we can see that by
employing the ABARO algorithm which optimizes the adjustable matrices after
each transmission contributes to a lower error probability upper bound.} As
shown in the previous section, by employing adjustable code matrices and the
proposed ABARO algorithm, an improvement of the coding gain is obtained which
reduces the error probability.

\begin{figure}
\begin{center}
\def\epsfsize#1#2{0.65\columnwidth}
\epsfbox{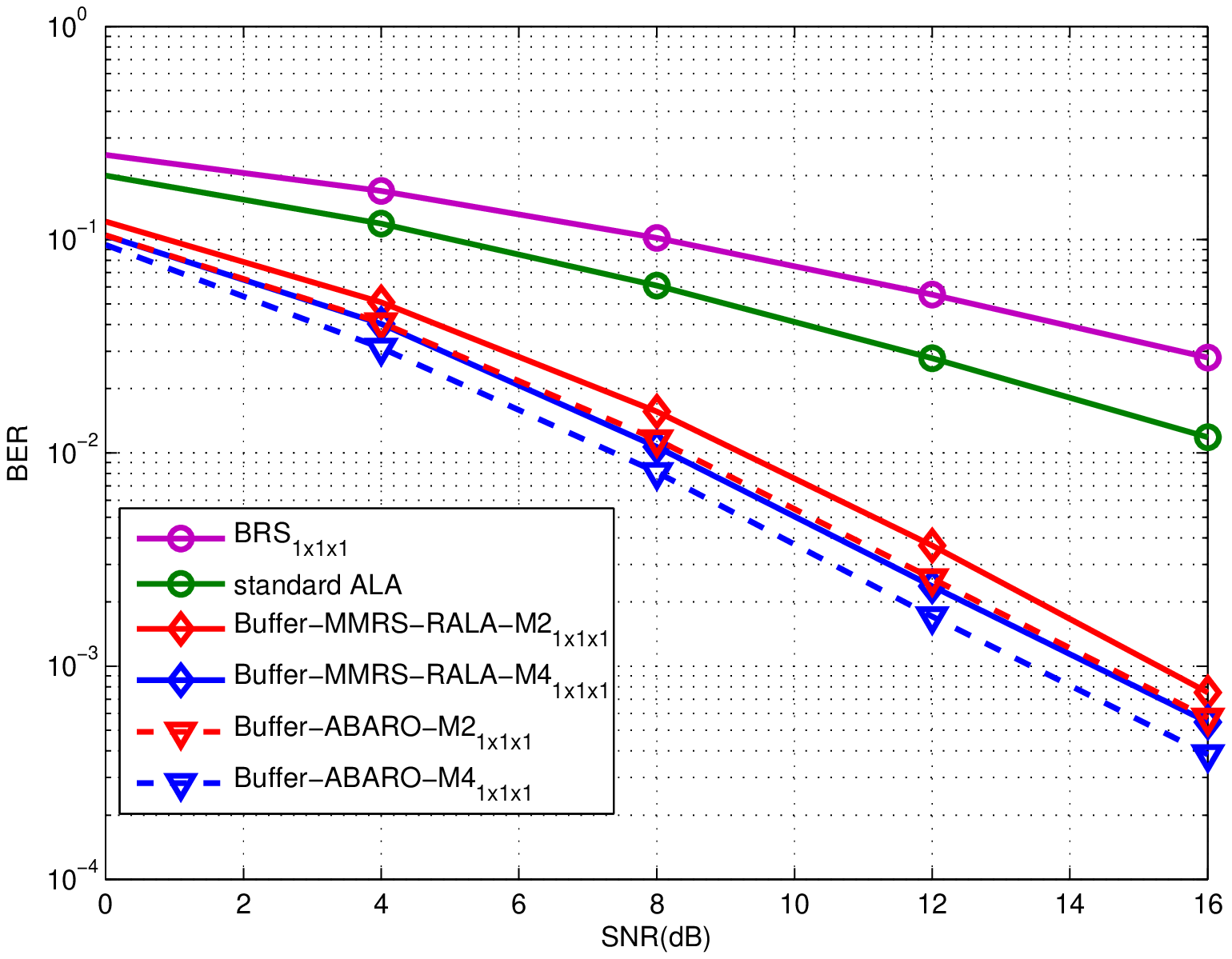}\vspace*{-1em} \caption{BER Performance vs.
$SNR$ for SAS, $1$ Relay}\label{f2}
\end{center}
\end{figure}

The proposed ABARO algorithm with the Alamouti scheme and an ML
receiver in the SAS configuration is evaluated with a single-relay
system in Figs. 4 and 5. Different buffer sizes are considered at
the relay node.  {A static channel is employed during the simulation
and the corresponding period in which the channel is static is equal
to one symbol.}  {The BER results of the cooperative system with the
best relay selection (BRS) algorithm in \cite{A.Bletsas} and the
max-max relay selection (MMRS) protocols in \cite{A.Ikhlef2} are
shown in both figures. The BER performance of using standard
Alamouti scheme at the relays is given as well. No STC schemes are
used in Fig. 5 so that the curves achieve a first order diversity. A
$2$dB to $3$dB BER improvement in Fig. 5 can be observed by
employing MMRS algorithms as compared to the BRS algorithm.
According to the simulation results, with the increase of the buffer
size at the relay nodes, the improvement in the BER reduces and with
the buffer size greater than $6$ the advantages of using
buffer-aided relays are not that obvious.}
\begin{figure}
\begin{center}
\def\epsfsize#1#2{0.65\columnwidth}
\epsfbox{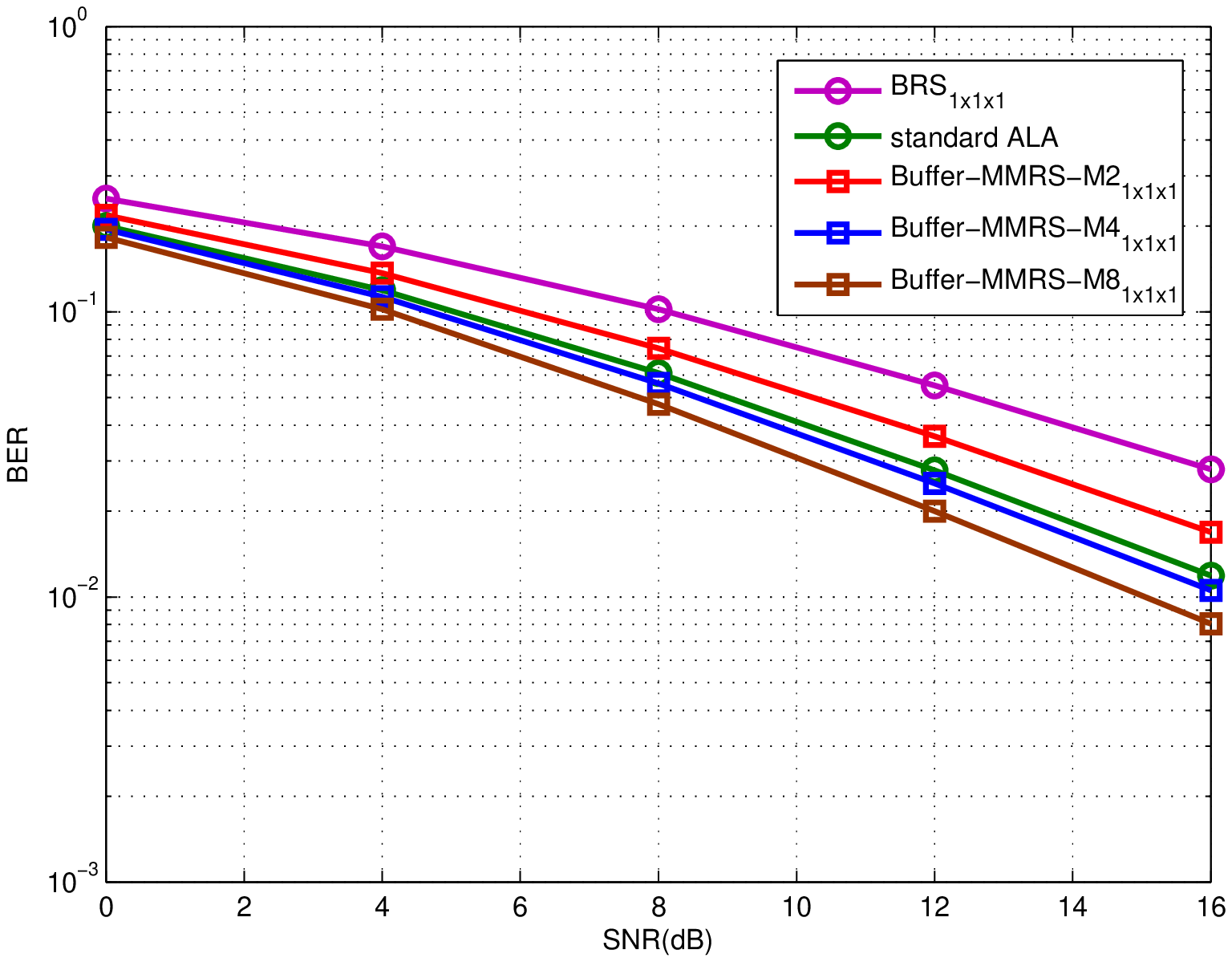}\vspace*{-1em} \caption{BER Performance vs.
$SNR$ for SAS, $1$ Relay}\label{f2}
\end{center}
\end{figure}

An improvement of diversity order can be observed when using STBC
schemes at the relays which is shown in Fig. 6. With the buffer size
greater than $4$, the advantage of using STBC schemes at the relays
disappears as a function of the diminishing returns in performance.
As shown in the simulation results, when the RSTC scheme is
considered at the relay node, the BER curve with buffer size of $6$
approaches that with buffer size of $8$ as well.  {In Fig. 6, the
proposed ABARO algorithm is employed in the single-antenna systems
with $n_r=2$ relay nodes. According to the simulation results in
Fig. 6, a $1$dB to $2$dB gain can be achieved by using the proposed
ABARO algorithm at the relays as compared to the network using the
RSTC scheme at the relay node.} The diversity order of the curves
associated with the proposed ABARO algorithm is the same as that of
using the RSTC scheme at the relay node.  {Compared to the MMRS
algorithm derived in \cite{A.Ikhlef2} with the same buffer size, the
ABARO algorithm achieves a $1$dB to $2$dB improvement.}

\begin{figure}
\begin{center}
\def\epsfsize#1#2{0.65\columnwidth}
\epsfbox{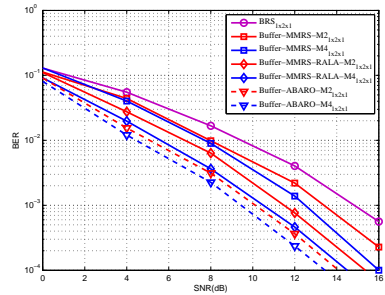}\vspace*{-1em} \caption{BER Performance vs.
$SNR$ for Buffer-Aided Relaying System, $2$ Relays}\label{f4}
\end{center}
\end{figure}

\begin{figure}
\begin{center}
\def\epsfsize#1#2{0.65\columnwidth}
\epsfbox{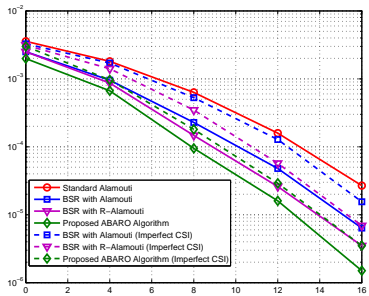}
\caption{BER Performance vs. $SNR$ for Buffer-Aided Relaying System}\label{f2}
\end{center}
\end{figure}

The proposed ABARO algorithm with the Alamouti scheme and an ML receiver is
evaluated in a MAS configuration with two relays in Fig. 7. It is shown in the
figure that the buffer-aided relay selection systems achieve $3$dB to $5$dB
gains compared to the previously reported relay systems. When the BSR algorithm
is considered at the relay node, an improvement of diversity order is shown in
Fig. 7 which leads to significantly improved BER performance. According to the
simulation results in Fig. 7, a $1$dB gain can be achieved by using the RSTC
scheme at the relays as compared to the network using the standard STC scheme
at the relay node. When the proposed ABARO algorithm is employed at the relays,
a $2$dB saving for the same BER performance as compared to the standard STC
encoded system can be observed. The diversity order of using the proposed ABARO
algorithm is the same as that of using the RSTC scheme at the relay node.

 {The impact of imperfect CSI at the destination node is
considered for different schemes as shown in Fig. 7. It is clear that a $2$dB
loss of the BER performance is obtained in BRS with Alamouti and R-Alamouti
schemes due to the imperfect CSI obtained at the destination node. As we
introduce errors in the channel elements in (\ref{2.1.9})-(\ref{2.1.12}), the
accuracy of optimal coding vector is affected. However, according to the
simulation result, a $1$dB BER loss is observed in Fig. 7 due to the channel
errors. The proposed optimization algorithm is able to maintain the BER
performance under the imperfect CSI obtained at the destination node. }

\section{Conclusion}

We have proposed a buffer-aided space-time coding scheme, relay selection and
the ABARO algorithms for cooperative systems with limited feedback using an ML
receiver at the destination node to achieve a better BER performance.
Simulation results have illustrated the advantage of using the adjustable STC
and DSTC schemes in the buffer-aided cooperative systems compared to the BRS
algorithms. In addition, the proposed ABARO algorithm can achieve a better
performance in terms of lower BER at the destination node as compared to prior
art. The ABARO algorithm can be used with different STC schemes and can also be
extended to cooperative systems with any number of antennas.

\bibliographystyle{IEEEbib}
\bibliography{strings,refs}

\bibliographystyle{IEEEtran}

\end{document}